\documentclass[12pt]{article}

\usepackage{a4wide}

\usepackage{bm}
\usepackage[dvipdfmx]{graphicx}
\usepackage{amsmath}
\usepackage{amsfonts}
\usepackage{amssymb}
\newcommand{\ol}{\overline}
\newcommand{\wt}{\widetilde}

\newcommand{\CC}{\mathbb{C}}
\newcommand{\ZZ}{\mathbb{Z}}
\newcommand{\RR}{\mathbb{R}}
\newcommand{\RP}{\mathbf{RP}}
\newcommand{\CP}{\mathbf{CP}}

\def\tr{\mathop{\rm tr}\nolimits}

\def\Puff{\mathop{\rm Puff}\nolimits}
\def\vol{\mathop{\rm Vol}\nolimits}

\newcommand{\rap}[2]
{\setbox1=\hbox{#1}%
\setbox2=\hbox to\wd1{\hss #2\hss}%
\mbox{\rlap{\box1}\box2}}

\makeatletter

\@addtoreset{equation}{section}
\makeatother

\def\half{{1\over2}}

\def\cN{{\cal N}}

\def\mbf{\mathbf}

\def\tr{\mathrm{tr}}

\def \bes#1 {\begin{equation}\begin{split}#1\end{split}\end{equation}}
\def \beal#1 {\begin{align}#1\end{align}}
\def \be {\begin{equation}}
\def \ee {\end{equation}}

\title{Superconformal index of ${\cal N}=3$ orientifold theories}

\begin{document}

\begin{titlepage}
\title{
\begin{flushright}
\normalsize{
TIT/HEP-652\\
March 2016}
\end{flushright}
       \vspace{2cm}
       Superconformal index of $\cN = 3$ orientifold theories
       \vspace{2cm}}
\author{
Yosuke Imamura\thanks{E-mail: \tt imamura@phys.titech.ac.jp}$^{~1}$, Shuichi Yokoyama\thanks{E-mail: \tt shuichitoissho@gmail.com}$^{~2}$
\\[30pt]
{\it $^1$ Department of Physics, Tokyo Institute of Technology,}\\
{\it Tokyo 152-8551, Japan}\\
{\it $^2$ Department of Physics,  Keio University,}\\
{\it Research and Education Center for Natural Sciences,}\\
{\it Kanagawa 223-8521, Japan}
}
\date{}

\thispagestyle{empty}

\vspace{0cm}

\maketitle
\begin{abstract}
We analyze the superconformal index
of the ${\cal N}=3$ supersymmetric
generalized orientifold theories recently proposed.
In the large $N$ limit we derive the index
from the Kaluza-Klein modes in $\bm{AdS}_5\times\bm{S}^5/\ZZ_k$,
which are obtained from ones in $\bm{AdS}_5\times\bm{S}^5$
by a simple $\ZZ_k$ projection.
For the ordinary $\ZZ_2$ orientifold case
the agreement with the gauge theory calculation
is explicitly confirmed, and for $\ZZ_{k\geq 3}$
we perform a few consistency checks with known results
for ${\cal N}=3$ theories.
We also study finite $N$ corrections
by analyzing wrapped D3-branes and discrete torsions
in the dual geometry.
\end{abstract}
\end{titlepage}


\section{Introduction}
Various four dimensional superconformal field theories
have been extensively studied for decades,
except one with just twelve supercharges.  
This is because the ${\cal N}=3$ vector multiplet is the only
free multiplet of ${\cal N}=3$ supersymmetry.
It has the same field contents as
those of the ${\cal N}=4$ vector multiplet, and thus
an ${\cal N}=3$ theory is necessarily enhanced
to the ${\cal N}=4$ one as long as it has perturbative description.
From the viewpoint of AdS/CFT correspondence \cite{Maldacena:1997re}, 
non-trivial $\cN=3$ superconformal field theories have been expected
to exist from analysis of type IIB supergravity solution \cite{Ferrara:1998zt}.
They have shown that an $\bm{AdS}_5$ solution with such superconformal symmetry
can be constructed by appropriate truncation from maximally supersymmetric one,
and also implied that some non-perturbative symmetry is mandatory to realize genuine ${\cal N}=3$ theory.
(See \cite{Beck:2016lwk} for recent study of such supergravity solution.)

A detailed realization of such genuin $\cN=3$ theories has recently been done in \cite{Garcia-Etxebarria:2015wns}.
They started with the D3-brane realization of the ${\cal N}=4$ $U(N)$ supersymmetric Yang-Mills theory (SYM),
and reduced its supersymmetry by
a simple orbifold projection at special points of the marginal coupling $\tau$.
For a generic value of $\tau$
a $\ZZ_2$ subgroup of the $SL(2,\ZZ)$ Montonen-Olive duality of the parent ${\cal N}=4$ theory
is a perturbative symmetry,
and the orbifold by it leads to the usual orientifold theories.
However,
the symmetry is enhanced to $\ZZ_k$ with $k=3,4,6$ at some special values of $\tau$,
and we can define new orbifolds,
which we refer to as $\ZZ_k$ orientifolds.
In the F-theoretic description,
the $\ZZ_k$ symmetry
can be interpreted as the rotation of the internal torus.

A general analysis of genuine ${\cal N}=3$ theories
that does not rely on explicit construction of the
theory was performed in \cite{Aharony:2015oyb}
by using ${\cal N}=3$ superconformal symmetry and its representations.
They proved that a genuine ${\cal N}=3$ theory
does not have exactly marginal deformations.
This is consistent with the fact that they can be defined only at special values of $\tau$.
This was also confirmed in \cite{Cordova:2016xhm},
in which a comprehensive analysis of supersymmetric deformations in superconformal theories was given.
Another important property is that
the dimension of Coulomb branch operators
must be integer greater than $2$.
This was confirmed in
rank-$1$ ${\cal N}=3$ theories in \cite{Argyres:2016xua,Nishinaka:2016hbw}.

A purpose of this paper is
to analyze the $\ZZ_k$ orientifold theories
defined in \cite{Garcia-Etxebarria:2015wns} from their holographic duals.
Namely, type IIB theory in $\bm{AdS}_5\times\bm{S}^5/\ZZ_k$ with $k=3,4,6$.
We first derive the superconformal index of such generalized orientifold theories
in the large $N$ limit by means of Kaluza-Klein analysis.
Fortunately, the $SL(2,\ZZ)$ symmetry
is realized as a classical symmetry
in type IIB supergravity,
which facilitates specification of the orientifold.
Since the Kaluza-Kline modes in $\bm{AdS}_5\times\bm{S}^5$ were already examined
in \cite{Gunaydin:1984fk}, it is straightforward to calculate
the index of those on the orientifold by a simple projection. 
We check that the index determined by the truncation precisely
reproduces that of the corresponding gauge theory when $k=2$,
in which case the gauge theory is $\cN=4$ SYM
with orthogonal or symplectic gauge group,
and its large $N$ index is computable.

Then we move on to analysis on finite $N$ corrections of the index.
In the perturbative cases with $k=1$ and $2$
the leading correction to the superconformal index
in the finite $N$ theory comes from wrapped D3-branes.
If they wrap topologically trivial cycles
they are called giant or dual-giant gravitons,
and give negative contribution to the index.
If the cycle is topologically non-trivial
they contribute positively to the index.
We assume this relation persists in general cases
and determine which type of wrapped branes gives the leading
correction to the index
by analyzing the discrete torsion
and the three-cycle homology.

\subsection*{Note added}
After completion of this work,
we have noticed a paper by Aharony and Tachikawa \cite{Aharony:2016kai},
which also analyze ${\cal N}=3$ theories from a holographic point of view
and has some overlap with this work.

\section{$\ZZ_k$ orientifolds}
\label{orientifold} 

The orientifold construction of
${\cal N}=3$ theories in \cite{Garcia-Etxebarria:2015wns}
starts from the type IIB realization of the ${\cal N}=4$
supersymmetric $U(N)$ Yang-Mills theory.
We realize it as the theory on a stack of $N$ D3-branes placed in
the flat ten-dimensional space-time.
It possesses
the superconformal symmetry
$PSU(2,2|4)$,
whose bosonic subgroup is
\begin{align}
SU(2,2)\times SU(4)_R
\subset PSU(2,2|4).
\end{align}
The $R$-symmetry, $SU(4)_R$, is the rotation of $\RR^6$
transverse to the D3-branes.
The four
supercharges $Q_I$ ($I=1,2,3,4$)
with positive chirality belong to
the fundamental representation of $SU(4)_R$.
We represent $SU(4)_R$ irreducible representations
by using integral Dynkin labels $(r_1,r_2,r_3)$.
The integers $r_a$ are eigenvalues of Cartan generators
$R_a$, whose action on $Q_I$ are
\begin{align}
R_1=(1,-1,0,0),\quad
R_2=(0,1,-1,0),\quad
R_3=(0,0,1,-1).
\end{align}
(We mean by $(\lambda_1,\lambda_2,\lambda_3,\lambda_4)$
that the eigenvalues for $Q_I$ are $\lambda_I$.)
In our convention the Dynkin labels of the fundamental and the anti-fundamental
representations are $(1,0,0)$ and $(0,0,1)$, respectively.
Unlike the ${\cal N}\leq 3$ case the $R$-symmetry group is
not $U({\cal N})_R$ but $SU(4)_R$.

For the conformal group, we use the Cartan generators
of the compact subgroup $U(1)\times SU(2)_1\times SU(2)_2\subset SU(2,2)$:
the conformal energy $E$,
the $SU(2)_1$ Cartan $J_1$, and the $SU(2)_2$ Cartan $J_2$.
We normalize these generators so that the supercharges $Q_I$ carry
$E=\frac{1}{2}$ and $J_{1/2}=0,\pm1$.

Let us consider an orientifold group
generated by $g$ that acts on the supercharges as
\begin{align}
g:
(Q_1,Q_2,Q_3,Q_4)\rightarrow
(e^{i\theta_1}Q_1,e^{i\theta_2}Q_2,e^{i\theta_3}Q_3,e^{i\theta_4}Q_4).
\label{gonq}
\end{align}
To preserve three supercharges $Q_I$ ($I=1,2,3$),
we need to set $\theta_1=\theta_2=\theta_3=0$.
When $g$ is an element of $SU(4)_R$
it is impossible to obtain a genuine ${\cal N}=3$ theory
because
$\theta_{1,2,3}=0$ requires the fourth angle $\theta_4$ to be also zero,
and the whole ${\cal N}=4$ supersymmetry is preserved.
To realize genuine ${\cal N}=3$ supersymmetry, we need to combine $SU(4)_R$
with the duality symmetry, which rotates all $Q_I$ by the same angle.

The $SL(2,\RR)$ symmetry of type IIB supergravity
can be linearly realized by introducing auxiliary scalar fields $V_a^+$ \cite{Schwarz:1983qr,Schwarz:1983wa}.
In such a formalism the theory
has the $SL(2,\RR)$ global and a $U(1)$ local symmetries,
and the scalar fields $V_a^+$ ($a=1,2$) are transformed as $\bm{2}_{+1}$ under
the symmetry.
The moduli space $SL(2,\RR)/U(1)$ is parameterized by the
axio-dilaton $\tau=\chi+i e^{-\phi}=V_2^+/V_1^+$,
which is $U(1)$ invariant and transformed under $SL(2,\RR)$ as
\begin{align}
\tau\rightarrow \tau'=\frac{a\tau +b}{c\tau+d},\quad
a,b,c,d\in\RR,\quad
ad-bc=1.
\end{align}
The expectation values of $V_a^+$ break $SL(2,\RR)\times U(1)$ down to
a $U(1)$ global symmetry, which is denoted in
\cite{Garcia-Etxebarria:2015wns} by $U(1)_Y$.
This plays an essential role
in the orientifold construction of ${\cal N}=3$ theories.
We denote its generator by $Y$.

The three-form fields $H_3^{\rm RR}$ and $H_3^{\rm NS}$,
and the corresponding $(p,q)$-string charges
\begin{align}
q^1=\frac{1}{2\pi}\int H_3^{\rm RR},\quad
q^2=\frac{1}{2\pi}\int H_3^{\rm NS},
\end{align}
form $SL(2,\RR)$ doublets.
Two charges are often combined into the complex charge
\begin{align}
q^C=V_a^+q^a.
\end{align}
The quantized values of $q^C$ form the two-dimensional charge lattice
with the modulus $\tau$ on the complex plane.
The associated torus is the internal torus of the
F-theory description of type IIB theory.
Because the $U(1)_Y$ symmetry rotates the lattice
the symmetry is broken to a discrete subgroup $\Gamma\subset U(1)$.
For a generic value of $\tau$ this is $\Gamma=\ZZ_2$,
while at special values of $\tau$ larger symmetry is
realized;
$\Gamma=\ZZ_k$ with $k=3,4,6$ for $\tau=e^{2\pi i/k}$ up to $SL(2,\ZZ)$.
The orientifold construction of genuine ${\cal N}=3$ theories
works only for these special values of $\tau$.

To describe $U(1)_Y$ transformation,
it is convenient to combine two three-form fields into a complex field
$H_3^C=V_1^+H_3^{\rm RR}+V_2^+H_3^{\rm NS}$.
We also express the spin $3/2$ and spin $1/2$ fermion fields
as complex fields
$\psi_\mu^C$ and $\lambda^C$, respectively.
If we normalize the $U(1)_Y$ generator $Y$ so that $Y(V_a^+)=+1$
the complex fields carry the following charges:
\begin{align}
Y(\psi_\mu^C)=+\frac{1}{2},\quad
Y(H_3^C)=+1,\quad
Y(\lambda^C)=+\frac{3}{2},\quad
Y(\delta\tau)=+2,
\end{align}
where $\delta\tau$ denotes the fluctuation around the expectation value
of $\tau$.
The non-trivial $U(1)_Y$ charge of the gravitino
implies that the supercharge is also rotated by
$U(1)_Y$.
In the context of the boundary ${\cal N}=4$ theory
all the four supercharges $Q_I$ ($I=1,2,3,4$)
carries $Y=+1/2$.

Let $z_i$ ($i=1,2,3$) be complex coordinates
of the transverse space $\RR^6=\CC^3$.
We consider the abelian orbifold group
that is generated by the combination of the spacial rotation
\begin{align}
(z_1,z_2,z_3)\rightarrow
(e^{i\varphi_1}z_1,e^{i\varphi_2}z_2,e^{i\varphi_3}z_3)
\end{align}
and the $U(1)_Y$ rotation $e^{i\varphi_4Y}$.
This acts on the supercharges as
the rotation (\ref{gonq}) with the angles
\bes{ 
\theta_1&=\frac{1}{2}(\varphi_1-\varphi_2-\varphi_3+\varphi_4), \\
\theta_2&=\frac{1}{2}(-\varphi_1+\varphi_2-\varphi_3+\varphi_4), \\
\theta_3&=\frac{1}{2}(-\varphi_1-\varphi_2+\varphi_3+\varphi_4), \\
\theta_4&=\frac{1}{2}(\varphi_1+\varphi_2+\varphi_3+\varphi_4).
}
We set the same value to all $\varphi_i$
\begin{align}
\varphi_1=\varphi_2=\varphi_3=\varphi_4=\varphi
\end{align}
and denote the corresponding rotation by $g=e^{i\varphi A}$,
where the generator $A$ acts on $Q_I$ as
\begin{align}
A=(0,0,0,+2).
\end{align}

As we mentioned the $U(1)_Y$ symmetry is broken
to its discrete subgroup,
and only restricted values of $\varphi$ is allowed.
To realize $\ZZ_k$ orientifold, we take
\begin{align}
\varphi=\frac{2\pi}{k},\quad
k=2,3,4,6.
\end{align}
The fourth supercharge $Q_4$ is rotated by $g$ as
\begin{align}
Q_4\rightarrow e^{\frac{4\pi i}{k}}Q_4,
\end{align}
which is trivial for $k=2$ and non-trivial for $k=3,4,6$.  
Accordingly while the orbifold projection for the former value of $k$ gives another $\cN=4$ theory,
the projections for the latter values give genuine ${\cal N}=3$ theories.

It is natural to ask what kind of gauge theory is
realized on the D3-branes  sitting
at the $\ZZ_k$ orientifold plane.
For $k=2$ the projection is
nothing but the ordinary perturbative orientifold
associated with the reversal of string orientation.
Leaving string modes invariant under the orientifold action
leads to the orthogonal or symplectic gauge group
depending on the intrinsic parity of open strings.
When the number of D3-branes before the projection
is $2N$ the gauge group is $SO(2N)$ or $Sp(N)$,
while $SO(2N+1)$ gauge theory is realized for $2N+1$ D3-branes.
For $\cN=3$ theories
this question is rather difficult to answer
due to the fact that
the theory is defined in a non-perturbative fashion
(see \cite{Aharony:2015oyb} for recent analysis).

Before ending this section,
let us explicitly define the superconformal index, which will be analyzed in the following sections.
To define the index, we should choose one of the supercharges.
We use the one with the following quantum numbers.
\begin{align}
&E(Q)=\frac{1}{2},\quad
J_1(Q)=-1,\quad
J_2(Q)=0,\nonumber\\
&R_1(Q)=1,\quad
R_2(Q)=R_3(Q)=0.
\end{align}
The choice of the supercharge breaks $SU(4)_R$ symmetry into $SU(3)\times U(1)$.
Furthermore, the orientifold projection with $k\geq3$ further breaks the $SU(3)$ into
$SU(2)\times U(1)$.
Thus it is convenient to define the following additional $SU(4)_R$ Cartan generators.
\begin{align}
X&=3R_1+2R_2+R_1=(+3,-1,-1,-1),\nonumber\\
T_8&=R_2+2R_2=(0,+1,+1,-2).
\end{align}
$X$ appears in the algebra
\begin{align}
\Delta:=\{Q,Q^\dagger\}=E-J_1-\frac{1}{2}X,
\end{align}
and the $SU(3)\times U(1)$ subgroup respected by the
choice of $Q$ is specified as the maximal subalgebra of $SU(4)_R$
commuting with $X$.
We use $T_8$ together with $R_2$ as Cartan generators of $SU(3)$,
and we define the ${\cal N}=4$ superconformal index \cite{Kinney:2005ej} by
\begin{align}
I(t,y,p,q)=\tr\left[(-1)^Fe^{-\beta\Delta}t^{2E+J_1}y^{J_2}p^{R_2}q^{T_8}\right].
\label{scin4}
\end{align}

\section{Large $N$ limit}
\label{gravityindex}

A $\ZZ_k$ orientifold theory is characterized by $k$, the order of the orientifold group,
and $N$, the size of the gauge group.
Precisely speaking, the latter does not make sense for $k=3,4,6$ because
of the absence of the perturbative description.
Even so we can define $N$
as the dimension of the Coulomb branch divided by $6$.
This is often referred to as the rank of the theory.
From the viewpoint of the brane system
the rank is the number of mobile D3-branes.
In this section we analyze the superconformal index in the large $N$ limit
by using the AdS/CFT correspondence.

In general, $k$ and $N$ do not uniquely specify the theory
because there may be more than one theories
distinguished by discrete torsions.
The difference among these theories
does not appear in the large $N$ limit.
We  discuss finite $N$ corrections in the
next section, and here we do not pay attention to the
discrete torsions.

The Kaluza-Klein spectrum of type IIB supergravity in
$\bm{AdS}_5\times\bm{S}^5$
was determined in \cite{Gunaydin:1984fk}.
They belong to the reducible representation
\begin{align}
\bigoplus_{n=1}^\infty {\cal S}[0;0]_{n,0}^{(0,n,0)}
\label{eq31}
\end{align}
of the superconformal algebra.
${\cal S}[j_1,j_2]_{E,Y}^{(r_1,r_2,r_3)}$ denotes the
superconformal irreducible representation
with highest weight state that carries the specified quantum numbers.
Each irreducible representation
in (\ref{eq31}) is
decomposed into $36$ irreducible representations of
the bosonic subgroup $SU(2,2)\times SU(4)$ for generic $n$,
which are listed in the table in \cite{Gunaydin:1984fk}.
\begin{align}
{\cal S}[0;0]_{n,0}^{(0,n,0)}
=&{\cal C}[0;0]_{n,0}^{(0,n,0)}
+{\cal C}[0;1]_{n+\frac{1}{2},-\frac{1}{2}}^{(1,n-1,0)}
+{\cal C}[0;0]_{n+1,-1}^{(2,n-2,0)}
\nonumber\\&
+{\cal C}[1;0]_{n+\frac{1}{2},+\frac{1}{2}}^{(0,n-1,1)}
+{\cal C}[1;1]_{n+1,0}^{(1,n-2,1)}
+{\cal C}[1;0]_{n+\frac{3}{2},-\frac{1}{2}}^{(2,n-3,1)}
\nonumber\\&
+{\cal C}[2;0]_{n+1,+1}^{(0,n-1,0)}
+{\cal C}[2;1]_{n+\frac{3}{2},+\frac{1}{2}}^{(1,n-2,0)}
+{\cal C}[2;0]_{n+2,0}^{(2,n-3,0)}
\nonumber\\&
+\cdots.
\label{to36}
\end{align}
We denote the $SU(2,2)\times SU(4)_4\times U(1)_Y$ irreducible representations
by ${\cal C}[j_1;j_2]^{(r_1,r_2,r_3)}_{E,Y}$.
Only nine of them in the decomposition
saturating the BPS bound $\Delta\geq0$ are shown explicitly in (\ref{to36}),
and $\cdots$ denote the other terms that do not contribute to the index.
For small $n$ some of representations on the right hand side are absent.
By using this decomposition
it is easy to calculate the superconformal index.
Let $S_n$ be the contribution of an ${\cal N}=4$ superconformal
multiplet ${\cal S}[0;0]_{n,0}^{(0,n,0)}$.
It is
given by
\begin{align}
S_n(t,y,p,q)
=\sum(-1)^F \frac{t^{2E+j_1}\chi_{j_2}(y)\chi_{(r_2,r_3)}(p,q)}{(1-t^3y)(1-t^3\tfrac{1}{y})}, 
\label{withoutz}
\end{align}
where the sum is taken over the nine components
explicitly shown in (\ref{to36}).
The $SU(2)$ character $\chi_s(y)$ and the $SU(3)$ character $\chi_{(m,n)}(p,q)$
are defined so that
\begin{align}
&\chi_s(y)=y^s+y^{s-2}+\cdots+y^{-s},\nonumber\\
&\chi_{(1,0)}(p,q)=q\left(p+\frac{1}{p}\right)+\frac{1}{q^2},\quad
\chi_{(0,1)}(p,q)=\frac{1}{q}\left(p+\frac{1}{p}\right)+q^2.
\end{align}
By summing up $S_n$ for $n=1,2,\ldots$,
we obtain the total single particle index \cite{Kinney:2005ej}.
\begin{align}
I^{\rm KK}(t,y,p,q)
&=\sum_{n=1}^\infty S_n(t,y,p,q).
\end{align}
An explicit expression is obtained by setting $z=1$ and $q'=q$
in (\ref{idepz}) below.

In the large $N$ limit, the index of the ${\cal N}=3$ theory is
obtained by the $\ZZ_k$ projection
that eliminate $\ZZ_k$ non-invariant modes.
For this purpose it is convenient to extend the index
by introducing the fugacity for $U(1)_Y$.
We combine $Y$ and $X$
to preserve $Q$ used for the definition of the index (\ref{scin4}),
and insert the factor
$z^{Y-\frac{X}{6}}$.
In addition,
we rewrite the fugacity $q$ by $q=q'z^{-2/3}$.
Then, by using the relation
\begin{align}
Y-\frac{1}{6}X=A+\frac{2}{3}T_8
\end{align}
we can rewrite the factor $q^{T_8}z^{Y-\tfrac{X}{6}}$
in the index as $q'^{T_8}z^A$.
As the result we obtain the index.
\begin{align}
&I^{\rm KK}(t,y,p,q',z)
=\tr\left[(-1)^Ft^{2E+J_1}y^{J_2}p^{R_2}q'^{T_8}z^A\right].
\end{align}

The $U(1)_Y$ charge of each Kaluza-Klein mode
is given in \cite{Gunaydin:1984fk},
and are shown in the decomposition (\ref{to36}).
It is straightforward
to repeat the calculation with the extra fugacity.
The contribution of
each superconformal multiplet is
\begin{align}
S_n(t,y,p,q',z)
&=\sum(-1)^F \frac{t^{2E+j_1}\chi_{j_2}(y)\chi_{(r_2,r_3)}(p,q'z^{-2/3})z^{Y-x/6}}
{(1-t^3y)(1-t^3\tfrac{1}{y})},
\label{withz}
\end{align}
where the summation is again taken over the nine components
explicitly shown in (\ref{to36}).
Although fractional powers of $z$ appear in the expression
(\ref{withz}) we can easily check that powers of $z$ in the $z$-expansion of $S_n$ are actually always $n$ mod $2$.
Namely, $S_n$ satisfies
\begin{align}
S_n(t,y,p,q',-z)
=(-1)^nS_n(t,y,p,q',z).
\label{zparity}
\end{align}

By summing up all the contributions for $n=1,2,\ldots$ we obtain
\begin{align}
&I^{\rm KK}(t,y,p,q',z)=\sum_{n=1}^\infty S_n(t,y,p,q',z)
\nonumber\\
&=\frac{(1-\tfrac{1}{z}t^3y)(1-\tfrac{1}{z}t^3\frac{1}{y})}{(1-t^3y)(1-t^3\frac{1}{y})}
\frac{1-t^4(\tfrac{z}{pq'}+\tfrac{zp}{q'}+\tfrac{q'^2}{z})+(1+z)t^6}
{(1-t^2\frac{pq'}{z})(1-t^2\frac{q'}{zp})(1-t^2\frac{z}{q'^2})}
\nonumber\\
&\hspace{2cm}-\frac{1-t^6\tfrac{1}{z}}{(1-t^3y)(1-t^3\frac{1}{y})}.
\label{idepz}
\end{align}
Once we obtain this expression, it is straightforward to perform
the projection.
We define $I_m^{\rm KK}$ by expanding
$I^{\rm KK}$ with respect to $z$ as
\begin{align}
I^{\rm KK}(t,y,p,q',z)
=
\sum_{m\in\ZZ}I_m^{\rm KK}(t,y,p,q')z^m.
\end{align}
Then the index of $\ZZ_k$ orientifold theory is obtained by
\begin{align}
I_{\ZZ_k}^{\rm KK}(t,y,p,q')=\sum_{m\in k\ZZ}I_m^{\rm KK}(t,y,p,q').
\end{align}
For reference we show the explicit form of the index for $p=q'=1$:
\beal{
I_{\ZZ_k}^{\rm KK}(t,y,1,1)=& {a_k(t) + b_k(t) \chi_1(y)  \over (1-t^3 y)(1-t^3 y^{-1})}.
}
The coefficients $a_k$ and $b_k$ are given in Table \ref{table:coeff},
including $a_1$ and $b_1$ for the unprojected index $I^{\rm KK}_{\ZZ_1}\equiv I^{\rm KK}$
\begin{table}[t]
\begin{center} 
\begin{tabular}{|c|c|c|}
\hline
$k$ & $a_k$ & $b_k$\\ 
\hline  \hline
$1$ & $\frac{t^2(3+2t^4+t^6)}{1-t^2}$
    & $-\frac{t^3(1+2t^2)}{1-t^2}$ \\
$2$ & $\frac{t^4(6-t^2+5t^4+t^6+t^8)}{(1-t^2)(1+t^2)^2}$
    & $-\frac{t^5(3+3t^2+5t^4+t^6)}{(1-t^2)(1+t^2)^3}$\\
$3$ & $\frac{t^4(2-t^2+t^4)}{1-t^2}$
    & $-\frac{t^5}{1-t^2}$ \\
$4$ & $\frac{t^4(2-2t^2+9t^4+t^6+9t^8+3t^{12}+t^{14}+t^{16})}{(1-t^2)(1+t^2+t^4+t^6)^2}$
    & $-\frac{t^5(1+5t^4+5t^6+7t^8+2t^{10}+3t^{12}+t^{14})}{(1-t^2)(1+t^2)^3(1+t^4)^2}$ \\
$6$ & $\frac{t^4(2-6t^2+9t^4-6t^6+6t^8-3t^{10}+2t^{12}-t^{14}+t^{16})}{(1-t^2)(1+t^6)^2}$
    & $-\frac{t^5(1-2t^2+2t^4+4t^8-3t^{10}+t^{12}+t^{14})}{(1-t^4)(1+t^6)^2}$ \\
\hline 
\end{tabular} 
\end{center}
\caption{The coefficients of the $\cN=3$ indices for each $k$.}
\label{table:coeff}
\end{table}

Some comments are in order.

The complete agreement of
the index $I^{\rm KK}$ before the projection
and the index of ${\cal N}=4$ $U(N)$ theory
is confirmed in \cite{Kinney:2005ej}.
The gauge theory index that should be compared to the one-particle index
of the Kaluza-Klein modes is not the full index
defined as the trace over all states in ${\bm S}^3$ Hilbert space
but
the plethystic logarithm of the full index.
In the following, we always mean by the index the plethystic logarism.

Note that the gauge group for $k=1$ is not $SU(N)$ but $U(N)$.
The diagonal $U(1)$ corresponds to $S_1$:
\begin{align}
S_1=
\frac{1}{z}\frac{q'(p+\frac{1}{p})t^2-(y+\frac{1}{y})t^3-q'^2t^4+t^6}{(1-t^3y)(1-t^3\frac{1}{y})}
+z\frac{\frac{1}{q'^2}t^2-\frac{1}{q'}(p+\frac{1}{p})t^4+t^6}{(1-t^3y)(1-t^3\frac{1}{y})}.
\end{align}
This consists of only $z^{\pm1}$ terms
and is always projected out for $k\geq2$.

When $k=2$ $S_{2n+1}$ are projected out due to the relation (\ref{zparity}),
and the index is given by
\begin{align}
I_{\ZZ_2}^{\rm KK}=\sum_{n=1}^\infty S_{2n}.
\end{align}
The projection does not divide $S_n$ into
two or more parts.
This fact is consistent with the fact that
$\ZZ_2$ projection is the usual orientifold projection
and it does not break the ${\cal N}=4$ supersymmetry.
This index precisely agrees with the large $N$ limit of the index of
${\cal N}=4$ $SO(N)$ and $Sp(N)$ gauge theories, which is calculated in Appendix \ref{soindex}.

Difference of the supersymmetry between $k=1,2$ and $k=3,4,6$
is also seen by looking at $S_2$, which corresponds to
the energy-momentum multiplet of ${\cal N}=4$ theory.
In \cite{Aharony:2015oyb} the branching of the ${\cal N}=4$
energy momentum multiplet into ${\cal N}=3$ multiplets
are examined.
The bottom components belonging to $\bm{20}$ of $SU(4)_R$ is
decomposed into $\bm{8}_0+\bm{6}_{-4}+\ol{\bm{6}}_{+4}$ of $SU(3)\times U(1)$.
This can be seen in the
$z$-expansion of $S_2$
\begin{align}
S_2=\frac{1}{z^2}\frac{3t^4-4t^5-t^6+2t^7}{(1-t^3)^2}
+\frac{2t^4-2t^5-4t^6+4t^7+2t^8-2t^9}{(1-t^3)^2}
+z^2\frac{t^4-2t^6+t^8}{(1-t^3)^2}.
\end{align}
(We set $y=p=q'=1$.)
The $z^0$ term, which is never projected out,
corresponds to the ${\cal N}=3$ energy-momentum multiplet,
while the terms containing $z^{\pm2}$
correspond to the multiplets
that include the exactly marginal deformation
and the fourth supersymmetry current.
The latter is projected out
when $k\geq 3$, and genuine ${\cal N}=3$ theories are realized.

For rank $1$ case a $\ZZ_k$ orientifold theory
has a Coulomb branch operator with dimension $k$
\cite{Nishinaka:2016hbw}.
This is expected to be the case for large $N$ limit, too, because
the $\ZZ_k$ invariant coordinates
of $\CC^3/\ZZ_k$ are given as order $k$ monomials
of the coordinates $z_i$ that have dimension $1$.
The corresponding contribution is found in the index.
For $\ZZ_k$ orientifold
$I^{\rm KK}_{\pm k}$ survive the projection
and when $k\geq 2$ the leading terms in $I_{\pm k}^{\rm KK}$ are given by
\begin{align}
I^{\rm KK}_{-k}&=t^{2k}\chi_k(p)q'^k+{\cal O}(t^{2k+1}),\nonumber\\
I^{\rm KK}_{+k}&=t^{2k}q'^{-2k}+{\cal O}(t^{2k+1}).
\end{align}
These are consistent with the existence of the ${\cal N}=3$ short representation
\begin{align}
{\cal S}[0;0]_k^{(0,k;2k)}\quad(\mbox{for $I^{\rm KK}_{-k}$}),\quad
{\cal S}[0;0]_k^{(k,0;-2k)}\quad(\mbox{for $I^{\rm KK}_{+k}$}),
\end{align}
where ${\cal S}[j_1;j_2]_E^{(r_1,r_2;r)}$ is
the ${\cal N}=3$ representation with
highest weights specified.
The quantum numbers $(r_1,r_2;r)$ are the $SU(3)_R$ integral Dynkin weights $(r_1,r_2)$ and
$U(1)_R$ charge $r$.
These are ${\cal N}=3$ chiral representations,
and their bottom components are identified with
the Coulomb branch operators with dimension $k$.

\section{Finite $N$ corrections}
\label{finiteN}

Now let us turn to finite $N$ corrections.
It is instructive to start from
the $k=1$ and $k=2$ cases
where we can calculate the index
on the gauge theory side.

For $k=1$ the theory is the ${\cal N}=4$ $U(N)$ SYM.
The index of the finite $N$ theory is related to $I^{\rm KK}=I_{U(\infty)}$ by
\begin{align}
I_{U(N)}&=I^{\rm KK}-\chi_{(N+1,0)}t^{2(N+1)}+{\cal O}(t^{2N+3}).
\label{uncorr}
\end{align}
(See (\ref{unindex}).)
The leading correction is of order $t^{2N+2}$
and its coefficient is negative.
On the gauge theory side this comes from the fact that
the gauge invariant operators of the form
\begin{align}
\tr\Phi^m
\label{trphim}
\end{align}
are independent only for $m=1,2,\ldots,N$,
and ones with $m\geq N+1$ can be decomposed into the independent ones.

On the gravity side,
they correspond to objects with angular momentum $\ell\sim m$.
When $\ell$ is of order $N$ they
should be treated not as point-like gravitons but as extenden objects.
A giant graviton \cite{McGreevy:2000cw} (a spherical D3-brane expanding in the $\bm{S}^5$)
and a dual giant graviton \cite{Grisaru:2000zn,Hashimoto:2000zp}
(a spherical D3-brane expanding in the $\bm{AdS}_5$) are two extremes of such objects.
The configuration relevant to the leading finite $N$ correction
to the index is
a giant graviton \cite{Balasubramanian:2001nh,Corley:2001zk,Corley:2002mj},
whose radius depending on $\ell$ is bounded by the radius of $\bm{S}^5$,
which is the same as the AdS radius
\begin{align}
L=(4\pi Ng_{\rm str})^{\frac{1}{4}}l_s.
\label{l4}
\end{align}
The mass $M_{\rm D3}$ of a D3-brane wrapped around a large $\bm{S}^3\subset\bm{S}^5$
in the unit $1/L$ is
\begin{align}
M_{\rm D3}L
=2\pi^2L^4T_{D3}
=N.
\end{align}
At the second equality we used $T_{\rm D3}=1/((2\pi)^3l_s^4g_{\rm str})$.
There are no giant gravitons whose energy exceeds this bound.
This is easily generalized to the $\ZZ_k$ orientifold.
Because a giant graviton wraps a topologically trivial cycle
we can use covering space $\bm{S}^5$ in the calculation of the bound,
and the only difference is that
the number of the flux $N$ is replaced by $kN$.

The $k=2$ case is investigated in \cite{Witten:1998xy} in detail.
An important feature of $k\geq2$ case
is that we have different choices of discrete torsion
classified by
\begin{align}
\Gamma_{\rm tor}^{(k)}=H^3(\bm{S}^5/\ZZ_k,\wt{\ZZ\oplus\ZZ}).
\label{tors3}
\end{align}
$\wt{\ZZ\oplus\ZZ}$ represents the sheaf of a pair of integers
corresponding to two three-form field strength
$H_3^{\rm RR}$ and $H_3^{\rm NS}$,
which has the $\ZZ_k$ monodromy
around the non-trivial cycle in $\bm{S}^5/\ZZ_k$.
When $k=2$
the internal space is $\RP^5=\bm{S}^5/\ZZ_2$
and the monodromy around the non-trivial cycle is
$(p,q)\rightarrow(-p,-q)$.
Because this monodromy does not mix the R-R and NS-NS fluxes
the torsion in the $k=2$ case is factorized into the direct sum
\begin{align}
\Gamma_{\rm tor}^{(k=2)}
=H^3(\RP^5,\wt\ZZ)\oplus H^3(\RP^5,\wt\ZZ)
=\ZZ_2\oplus\ZZ_2.
\end{align}
There are four cases, which
we denote by $(0,0)$, $(0,1)$, $(1,0)$ and $(1,1)$
where the first and the second components represent the discrete torsion
of $H_3^{\rm RR}$ and $H_3^{\rm NS}$, respectively.
They correspond to the following gauge groups \cite{Witten:1998xy}:
\begin{align}
(0,0):SO(2N),\quad
(1,0):SO(2N+1),\quad
(0,1):Sp(N),\quad
(1,1):Sp(N).
\end{align}
The latter three are transformed among them by the $SL(2,\ZZ)$
Montonen-Olive duality, while
the $SO(2N)$ theory corresponding to the trivial
discrete torsion is self-dual.

Let us first consider the $SO(2N)$ theory
corresponding to the trivial discrete torsion.
The index is
\footnote{The gauge group of the theory is actually $O(2N)$ rather than $SO(2N)$.
We treat $\ZZ_2=O(2N)/SO(2N)$ as a global symmetry
and we assume the trivial holonomy in the index calculation.}
\begin{align}
I_{SO(2N)}&=I^{\rm KK}_{\ZZ_2}+\chi_{(N,0)}t^{2N}+{\cal O}(t^{2N+1}).
\label{isoen}
\end{align}
(See (\ref{so2nindex}).)
We have the leading correction of order $t^{2N}$
with the positive coefficient.
On the gauge theory side
this is identified with the contribution of baryonic operators of the form
\begin{align}
B=\Puff\Phi
\label{fuffop}
\end{align}
with dimension $E=N$.
On the gravity side, these correspond to D3-branes
wrapped on the topologically non-trivial cycle.
Let $H^{(k)}$ be the $3$-cycle homology of the internal space $\bm{S}^5/\ZZ_k$.
When $k=2$ it is
\begin{align}
H^{(k=2)}=H_3(\RP^5,\ZZ)=\ZZ_2.
\end{align}
The non-trivial element of this homology group
is $\RP^3\subset \RP^5$.
As we have shown above the mass of a D3-brane wrapped on a large $\bm{S}^3$
is the same as the number of flux passing through ${\bm S}^5$ in the unit of
$1/L$, and it is $2N$ in the $k=2$ case.
By taking account of $\vol(\RP^3)=\vol(\bm{S}^3)/2$
we obtain the dimension $E\sim N$, which is consistent with the
correction in (\ref{isoen}).

When the discrete torsion is non-trivial,
the corresponding $SO(2N+1)$ theory (or $Sp(N)$ theory)
has the index
\begin{align}
I_{SO(2N+1)}=I_{Sp(N)}
&=I^{\rm KK}_{\ZZ_2}-(\chi_{(2N+2,0)}+\cdots)t^{4N+4}+{\cal O}(t^{4N+5}).
\label{so2np1}
\end{align}
(See (\ref{so2np1index}).)
This time we have the leading correction of order $t^{4N+4}$
with the negative coefficient,
which corresponds to operators with dimension $E\sim 2N$.
(We are interested in the ${\cal O}(N)$ behavior
of the dimension, and do not pay attention to ${\cal O}(1)$ shift.)
This is explained in the
same way as the $k=1$ case by the bound for the giant graviton.
What is important here is that 
the positive contribution of
baryonic operator of dimension $E\sim N$,
which gives the leading correction in the $SO(2N)$ case,
is absent.
In the context of the gauge theory this is simply because
we cannot define operators in the form (\ref{fuffop}).
On the gravity side
this is
explained by the topological obstruction
to wrapped D3-branes due to the non-vanishing discrete torsion
\cite{Witten:1998xy}.

Let $K^{(k)}$ be the charge lattice of
the electric and magnetic charges on a D3-brane wrapped on
the cycle corresponding to the generating element of
$H^{(k)}$, which has the topology $\bm{S}^3/\ZZ_k$.
\begin{align}
K^{(k)}\equiv H_0(\bm{S}^3/\ZZ_k,\wt{\ZZ\oplus\ZZ}).
\end{align}
When $k=2$ it is given by
\begin{align}
K^{(k=2)}
=H_0(\RP^3,\wt{\ZZ\oplus\ZZ})
=H_0(\RP^3,\wt{\ZZ}\oplus H_0(\RP^3,\wt{\ZZ})
=\ZZ_2\oplus\ZZ_2.
\end{align}
The three form fluxes, which are classified by $\Gamma_{\rm tor}^{(k)}$,
induce electric and magnetic charges
on a wrapped D3-brane.
This is described by the natural pairing
\begin{align}
f: H^{(k)}\times \Gamma^{(k)}_{\rm tor}\rightarrow K^{(k)}.
\end{align}
The flux conservation and the compactness of the wrapped D3-brane require
the pairing to vanish.
For the discrete torsion
$(1,0)$, $(0,1)$, and $(1,1)$,
wrapped D3-branes give non-trivial element of $K^{(k=2)}$
and such wrappings are prohibited.

In the following we generalize the argument above
to $k\geq 3$ cases.
For a general $k$ the $3$-cycle homology classifying wrapped D3-branes
is determined by using the Gysin exact sequence as
\begin{align}
H^{(k)}\equiv H_3(\bm{S}^5/\ZZ_k,\ZZ)=\ZZ_k.
\end{align}
We can obtain $K^{(k)}$ from the lattice $\ZZ\oplus\ZZ$ of electric and magnetic charges
by taking account of
the non-trivial monodromy around the non-trivial cycle in $\bm{S}^3/\ZZ_k$.
We identify elements in $\ZZ\oplus\ZZ$
related by the $\ZZ_k$ rotation.
As the result we obtain the following reduced charge lattice
\begin{align}
K^{(3)}=\ZZ_3,\quad
K^{(4)}=\ZZ_2,\quad
K^{(6)}=0.
\label{k346}
\end{align}

To analyze the discrete torsion (\ref{tors3}),
it may be convenient to visualize the elements of $\Gamma^{(k)}_{\rm tor}$
as their Poincar\'e duals,
which are classified by the two-cycle homology group
\begin{align}
\Gamma_{\rm tor}^{(k)}\cong H_2({\bm S}^5/\ZZ_k,\wt{\ZZ\oplus\ZZ}).
\label{tors2}
\end{align}
For $k\geq 3$ the orientifold group mixes the NS-NS and R-R three-form fluxes,
and the torsion does not factorize.
An element
of this group is
regarded as a string worldsheet $\Sigma$ carrying the NS-NS and R-R charges.
To visualize $\Sigma$
it is convenient to describe $\bm{S}^5/\ZZ_k$
as a Hopf fibration over $\CP^2$.
The Chern class of the fiber is $k\in H^2(\CP^2,\ZZ)=\ZZ$.
Let $B\sim\CP^1$ be a non-trivial two-cycle in $\CP^2$.
The restriction of the Hopf fibration also has
the Chern class $k$,
and has the topology $\bm{S}^3/\ZZ_k$.
By using the Mayer-Vietoris exact sequence
we can show the isomorphism
\begin{align}
\Gamma_{\rm tor}^{(k)}=H_2(\bm{S}^5/\ZZ_k,\wt{\ZZ\oplus\ZZ})\cong
H_2(\bm{S}^3/\ZZ_k,\wt{\ZZ\oplus\ZZ}),
\end{align}
and we can focus on the subspace $\bm{S}^3/\ZZ_k\subset\bm{S}^5/\ZZ_k$.

Roughly speaking,
a non-trivial element of $\Gamma_{\rm tor}^{(k)}$
can be represented as a worldsheet $\Sigma$
wrapped on a section over $B$.
Of course, due to the non-vanishing Chern class,
there is no global section over $B$.
Let $B'$ be the punctured sphere obtained
by removing a point $p$ from $B$,
and $C_p$ the fiber over $p$.
We can define a section $\Sigma'$ over $B'$,
which has the topology of a punctured sphere as $B'$.
The reason why we cannot fill in the puncture of $\Sigma'$
to obtain a global section over $B$
is that
the boundary of $\Sigma'$
is not just a point-like puncture but ${\bm S}^1$
winding around $C_p$.

In fact, we can remove this puncture to
obtain $\Sigma$ for $k\geq2$.
Let us first consider $k=2$ case.
The boundary of $\Sigma'$ winds $C_p$ twice.
This means that
two parts of the boundary of $\Sigma'$ meet along $C_p$.
Due to the non-trivial monodromy of the string charges
around $C_p$,
the two worldsheets have opposite string charge to each other,
and the boundary can be consistently covered by
a M\"obius strip
wrapped around $C_p$.
As the result we obtain $\Sigma=\RP^2$,
which wraps the non-trivial cycle
in $\RP^5$.

A generalization to $k\geq 3$ is straightforward.
The Chern class of the fibration over $B$ is
now $k$, and the boundary of the worldsheet $\Sigma'$ winds $k$ times around
$C_p$.
Namely, $k$ worldsheets,
which are different parts of $\Sigma'$, meet along $C_p$.
We can stitch them consistently
into a $k$-valent junction
because their $(p,q)$ string charges are
related by the $\ZZ_k$ monodromy and
sum up to zero.

Once we fix $\Sigma$ for $k\geq2$, an element of the discrete torsion is
specified by the $(p,q)$ string charge of the worldsheet.
If we move $\Sigma$ around the fibers its charges get
transformed by the monodromy
and thus
charges transformed to one another by $\ZZ_k$
must be identified.
After this identification,
we obtain the same groups as $K^{(k)}$ in (\ref{k346}).
This is also directly shown from the isomorphism
$H_i^{({\rm tor})}\cong H_{n-i-1}^{({\rm tor})}$
for an $n$-dimensional manifold.

By using the worldsheet representation of the discrete torsion,
we can easily determine the pairing $f:H\times \Gamma_{\rm tor}\rightarrow K$
as the intersection of the worldsheet and the D3-brane worldvolume.
Let us consider case by case.
\begin{itemize}
\item $k=1$

$k=1$ gives ${\cal N}=4$ $U(N)$ SYM.
The $3$-cycle homology is trivial and
there exist no baryonic operators.
The leading term of the finite $N$ correction is
due to the bound of giant gravitons at $E\sim N$.

\item $k=2$

$k=2$ gives the perturbative orientifold theories.
The $3$-cycle homology is $H^{(k=2)}=\ZZ_2$ and
the discrete torsion group and the charge lattice is
$\Gamma_{\rm tor}^{(k=2)}=K^{(k=2)}=\ZZ_2\oplus\ZZ_2$.
The pairing is
\begin{align}
f:\ZZ_2\times(\ZZ_2\oplus\ZZ_2)\rightarrow (\ZZ_2\oplus\ZZ_2).
\end{align}
This vanishes for a wrapped D3-brane
only when the discrete torsion is trivial.
Then we have positive correction at $E\sim N$.
Otherwise, the leading correction is negative one at $E\sim 2N$.

\item $k=3$

$H^{(k=3)}=\Gamma_{\rm tor}^{(k=3)}=K^{(k=3)}=\ZZ_3$.
The pairing is
\begin{align}
f:\ZZ_3\times \ZZ_3\rightarrow\ZZ_3,
\end{align}
and this vanishes for non-trivial $3$-cycle only when the discrete torsion
vanishes.
Then we have the positive correction at $E\sim N$.
Otherwise, we have the negative contribution at $E\sim 3N$.

\item $k=4$

$H^{(k=4)}=\ZZ_4$ and $\Gamma^{(k=4)}_{\rm tor}=K^{(k=4)}=\ZZ_2$.
The pairing is
\begin{align}
f:\ZZ_4\times\ZZ_2\rightarrow \ZZ_2.
\end{align}
When the discrete torsion is trivial
any winding number of a D3-brane is allowed,
and the leading correction is positive one at $E\sim N$.
Even when the discrete torsion is non-trivial,
we still have wrapped D3-branes with even wrapping number,
and we have positive correction at $E\sim 2N$.%
\footnote{The analysis here is based on the assumption that
when the pairing vanish we can wrap a D3-brane consistently.
However, there are some arguments that this naive expectation
may not be correct.
In \cite{Aharony:2016kai}
it is pointed out that in the $k=4$ case non-trivial wrapping
seems to be allowed only when
the discrete torsion is trivial.
The authors would like to thank Y.~Tachikawa
for notifying of this point.
}

\item $k=6$

$H^{(k=6)}=\ZZ_6$ and $\Gamma^{(k=6)}_{\rm tor}=K^{(k=6)}=0$.
Because of the absence of the non-trivial discrete torsion,
any winding of D3-branes is allowed,
and we have positive correction at $E\sim N$.

\end{itemize}

We summarize these results in Table \ref{corrtable}.
\begin{table}[htb]
\centering
\begin{tabular}{cccccc}
\hline
\hline
      &$H^{(k)}$& $K^{(k)}=\Gamma_{\rm tor}^{(k)}$ & torsion & sign & dim \\
\hline
$k=1$ &$0$& $0$        &  $0$      &  $-$ & $\sim N$ \\
\hline
$k=2$ &$\ZZ_2$& $\ZZ_2\oplus\ZZ_2$ & $(0,0)$ & $+$ & $\sim N$ \\
      &&                    & $(1,0)$, $(0,1)$, $(1,1)$ & $-$ & $\sim 2N$ \\
\hline
$k=3$ &$\ZZ_3$& $\ZZ_3$ & $0$ & $+$ & $\sim N$ \\
      &&         & $1$, $2$ & $-$ & $\sim 3N$ \\
\hline
$k=4$ &$\ZZ_4$& $\ZZ_2$ & $0$ & $+$ & $\sim N$ \\
      &&         & $1$ & $+$ & $\sim 2N$ \\
\hline
$k=6$ &$\ZZ_6$& $0$ & $0$ & $+$ & $\sim N$ \\
\hline
\end{tabular}
\caption{Finite $N$ corrections to the index.
The negative and positive signs show that the leading correction comes
from the bound on the giant gravitons and D3-branes wrapped on non-trivial
cycles, respectively.
The right-most column shows the ${\cal O}(N)$ part of the dimension of
operators that give the leading corrections.
}\label{corrtable}
\end{table}

\section{Conclusions and Discussion}
In this paper we have investigated the superconformal index
of ${\cal N}=3$
$\ZZ_k$ orientifold
superconformal theories
by using AdS/CFT correspondence.
The result in the large $N$ limit has been obtained by
a simple $\ZZ_k$ projection
of the Kaluza-Klein modes.
We have checked that the result is consistent with some known properties of
genuine ${\cal N}=3$ theories.
We have also discussed finite $N$ corrections
based on the assumption that
corrections with positive and negative coefficients
originate from
D3-branes wrapped on non-trivial and trivial cycles, respectively.
We have determined the signature
and large $N$ behavior of the exponent
of the leading term in the $t$-expansion
by examining the discrete torsion
of the three-form fluxes.

It would be interesting to investigate more on BPS operators of $\cN=3$ theories predicted from the index determined in this paper.  
When $k=2$, the microscopic origin of such BPS operators lies in BPS strings ending on D3-branes at the orbifold singularity. 
Thus it may be natural to expect that BPS operators in $\cN=3$ theory originate junctions connecting D3-branes.

In this paper we treated the wrapped branes classically,
and obtained only the coefficients of ${\cal O}(N^1)$ term
in the exponent.
To determine the exponent
including
${\cal O}(1)$ part
and $SU(3)_R$ representation
we need to perform quantum mechanical analysis of
the collective motion of the wrapped branes. 
It would be interesting to determine the configuration of such wrapped branes.  

We represent the discrete torsion as
two-cycles that are Poincar\'e dual to
the three-form flux,
and study how the wrapped branes are affected by
the discrete torsion.
The same two-cycles can be used to
realize domain walls.
We can wrap $(p,q)$ five branes
on the cycles.
They divide the four-dimensional
spacetime into two parts.
On the two sides of a domain wall
the discrete torsions are different
by the amount depending on the homology class
of the wrapped fivebrane.
It may be interesting to study
properties of
such domain walls
by using the brane realization mentioned above.

We studied the discrete torsion in type IIB picture.
In \cite{Garcia-Etxebarria:2015wns} discrete torsion in the M-theory dual
was investigated, and more variety than those of
type IIB picture were found.
It may be interesting to study the relation between discrete torsions
in type IIB theory and M-theory.
It may play important role when we discuss
${\bm S}^1$ compactification of the
${\cal N}=3$ theory to three-dimensional theories.

As another example of wrapped branes on torsion-cycles in the dual internal geometry, one can consider a wrapped D1-brane on $H_1(\mbf S^5/\ZZ_k)$.  
This object will correspond to a monopole with fractional charge in the field theory side. It would be interesting to investigate the properties of monopole with fractional charge from a purely field-theoretical viewpoint, as done for BPS states \cite{Grant:2008sk,Yokoyama:2014qwa}.  

We hope to come back to these issues in near future.

\section*{Acknowledgements}
We would like to thank S.~Kanno and H.~Kato for valuable discussions.
We would also like to thank T.~Okuda for helpful comments. 
The work of YI is partially supported by Grand-in-Aid for Scientific Research (C) (No.15K05044),
Ministry of Education, Science and Culture, Japan.
The work of SY is supported by the MEXT-Supported Program for the Strategic Research Foundation at Private Universities “Topological Science” (Grant No. S1511006).

\appendix 
\section{The indices for small $N$}
In this appendix we summarize the indices of
${\cal N}=4$ theories with small $N$,
which are used in the main text.

A general formula for the total index is  \cite{Kinney:2005ej}
\be
\mbf I^{\cN=4} = \int [dU] \exp\left[\sum_{m=1}^\infty \frac{1}{m}f(t^m, y^m, p^m, q^m) \chi_{\rm adj}(U^m)\right]
\label{gaugeindex}
\ee
where $f$ is the index for the {\it letters} of the $\cN=4$ $U(1)$ gauge theory,
which are the elementary BPS operators of the theory, given by 
\beal{
f(t,y,p,q) = {t^2\chi_{(1,0)}-t^3\chi_1 - t^4\chi_{(0,1)} + 2t^6 \over (1 -t^3y)(1-{t^3 \over y}) },
\label{letterindex}
}
where $\chi_{(m,n)}\equiv\chi_{(m,n)}(p,q)$ and $\chi_s\equiv\chi_s(y)$ are
the $SU(3)$ and the $SU(2)$ characters, respectively.
$[dU]$ is the Haar measure of the gauge group and $\chi_{\rm adj}(U^m)$
is the character for the adjoint representation thereof, which describe the contributions of holonomy.
The letter index \eqref{letterindex} contains both contributions
from a ${\cal N}=1$ vector multiplet and three ${\cal N}=1$ chiral ones,
whose zero point contributions to the index cancel.
Note also that the holonomy contribution is already factored out from the letter index
\eqref{letterindex} since all the fields belong to the adjoint representation. See also \cite{Dolan:2008qi,Spiridonov:2010qv}.

For $U(N)$ gauge group, $\chi_{\rm adj}$ and $\int[dU]$ are given by
\begin{align}
\chi_{\rm adj}(U) =& N+\sum_{m\not=n}e^{i(\alpha_m-\alpha_n)},\\
\int[dU] =& \frac{1}{N!}\prod_{m=1}^N \int_0^{2\pi}\frac{d\alpha_m}{2\pi} \prod_{n\not=m}(1-e^{i(\alpha_m-\alpha_n)}),
\end{align}
where $\alpha_m$ ($m=1,\ldots, N$) is the $m$-th holonomy.
It is straightforward to obtain the index in the form of $t$-expansion by direct calculation.
For $U(N)$ gauge groups with $N=1,2,3,4$ we obtain
\bes{
I_{U(1)}
=&I^{\rm KK}
-\chi_{2,0}t^4
+\chi_1\chi_{1,0}t^5
+(-\chi_{3,0}+\chi_{1,1}+1)t^6
-\chi_1\chi_{0,1}t^7
+{\cal O}(t^8),\\
I_{U(2)}
=&I^{\rm KK}-\chi_{3,0}t^6+\chi_1\chi_{2,0}t^7+(-\chi_{4,0}+\chi_{2,1}-\chi_{0,2}-\chi_{1,0})t^8
+{\cal O}(t^9),\\
I_{U(3)}
=&I^{\rm KK}-\chi_{4,0}t^8+{\cal O}(t^9),\\
I_{U(4)}
=&I^{\rm KK}-\chi_{5,0}t^{10}+{\cal O}(t^{11}).
\label{unindex}
}
These results show the pattern in (\ref{uncorr}).

For $SO(2N)$ gauge group the adjoint character and the Haar measure are given by 
\bes{
\chi_{\rm adj}(U) =& N+\sum_{m<n}\left(e^{i(\alpha_m+\alpha_n)}+e^{-i(\alpha_m+\alpha_n)}\right)
                        +\sum_{m\neq n}e^{i(\alpha_m-\alpha_n)},\\
\int[dU] =& \frac{1}{2^{N-1}N!}\prod_{m=1}^N\int_0^{2\pi}\frac{d\alpha_m}{2\pi}
                      \prod_{m<n}(1-e^{i(\alpha_m+\alpha_n)})(1-e^{-i(\alpha_m+\alpha_n)})
                        \prod_{m\neq n}(1-e^{i(\alpha_m-\alpha_n)}),
}
and the indices for small $N$ are
\begin{align}
I_{SO(2)}
&=I^{\rm KK}_{\ZZ_2}+\chi_{(1,0)}t^2-\chi_1t^3-(\chi_{(2,0)}+\chi_{(0,1)})t^4+2\chi_1\chi_{(1,0)}t^5+{\cal O}(t^6),\nonumber\\
I_{SO(4)}
&=I^{\rm KK}_{\ZZ_2}+\chi_{(2,0)}t^4-\chi_1\chi_{(1,0)}t^5+(-\chi_{1,1}+1)t^6
+\chi_1(\chi_{(2,0)}+\chi_{(0,1)})t^7+{\cal O}(t^8),\nonumber\\
I_{SO(6)}
&=I^{\rm KK}_{\ZZ_2}+\chi_{(3,0)}t^6-\chi_1\chi_{(2,0)}t^7+(\chi_{(1,0)}-\chi_{(2,1)})t^8+{\cal O}(t^9).
\label{so2nindex}
\end{align}
We see the pattern of (\ref{isoen}).

For $SO(2N+1)$ gauge group
\bes{
\chi_{\rm adj}(U) =& N+\sum_{m<n}\left(e^{i(\alpha_m+\alpha_n)}+e^{-i(\alpha_m+\alpha_n)}\right)
                        +\sum_{m\neq n}e^{i(\alpha_m-\alpha_n)}
                        +\sum_{m=1}^N(e^{i\alpha_m}+e^{-i\alpha_m}),\\
\int[dU] =& \frac{1}{2^NN!}\prod_{m=1}^N\int_0^{2\pi}\frac{d\alpha_m}{2\pi}
                      \prod_{m<n}(1-e^{i(\alpha_m+\alpha_n)})(1-e^{-i(\alpha_m+\alpha_n)})
                        \prod_{m\neq n}(1-e^{i(\alpha_m-\alpha_n)})\nonumber\\
&\hspace{5cm}                        \times\prod_{m=1}^N(1-e^{i\alpha_m})(1-e^{-i\alpha_m}),
}
and the indices for small $N$ are
\bes{
I_{SO(1)}
&=I^{\rm KK}_{\ZZ_2}-\chi_{2,0}t^4+\chi_1\chi_{1,0}t^5+{\cal O}(t^6),\\
I_{SO(3)}
&=I^{\rm KK}_{\ZZ_2}-(\chi_{4,0}+\chi_{0,2})t^8+\chi_1(\chi_{3,0}+\chi_{1,1})t^9+{\cal O}(t^{10}),\\
I_{SO(5)}
&=I^{\rm KK}_{\ZZ_2}-(\chi_{6,0}+\chi_{2,2}+1)t^{12}+{\cal O}(t^{13}),
\label{so2np1index}
}
and we find the pattern in (\ref{so2np1}).

\section{Large $N$ limit of $SO(N)$ index}  
\label{soindex}

In this appendix we give a brief derivation of large $N$ limit of an index for $\cN=4$ SYM with orthogonal and symplectic gauge groups. 
For this purpose it is sufficient to consider the case when the gauge group is $SO(2N)$, since the large $N$ index of $SO(2N)$ matches those of $SO(2N+1)$ and $Sp(N)$. 
In this appendix we use the same convention as used in \cite{Kinney:2005ej}. 

For this end we first restrict the range of holonomy from $0$ to $\pi$ by using the symmetry of the index under flip of sign of each holonomy for convenience.   
Then under the large $N$ limit the holonomy distributes densely between $0$ and $\pi$, whose density function we denote by $\rho$:
\be 
\rho(\alpha) = {1\over N} \sum_{m=1}^N \delta(\alpha-\alpha_m),
\ee
whose normalization is 
\be 
\int_0^\pi \rho(\alpha) = 1.
\ee
Since the range of $\alpha$ is restricted from $0$ to $\pi$, the density function can be expanded by cosine functions so that 
\be 
\rho(\alpha) = {1\over2\pi}\rho_0 + {1\over \pi} \sum_{k=1}^\infty \rho_k \cos(k\alpha)
\ee
where $\rho_k$ is given by 
\be 
\rho_k = 2 \int_0^\pi \rho(\alpha) \cos(-k\alpha).
\ee
By using the Fourier components of the density function, the large $N$ form of the Haar measure can be written as 
\be 
[dU] \to \prod_{k=1}^\infty d\rho_k \exp^{ - N^2 \sum_{k=1}^\infty {1\over 2k} \rho_{k}^2 + N \sum_{k=1}^\infty {1\over 2k} \rho_{2k} }
\ee
up to an overall constant, 
and the character is of a form such that 
\be 
\chi_{\rm adj}(U^m) \to  N^2 \sum_{k=1}^\infty {1\over 2k} \rho_{k}^2 - N \sum_{k=1}^\infty {1\over 2k} \rho_{2k}. 
\ee
Plugging these into the gauge index \eqref{gaugeindex} and performing the integrations by Gaussian ones we obtain the large $N$ index as 
\be 
\mbf I^{\cN=4}_{SO(\infty)} \sim \prod_{m \geq1} { e^{\frac{(1 - f(t^m, y^m, v^m, w^m))^2}{4m (1 - f (t^{2m}, y^{2m}, v^{2m}, w^{2m}))}} \over \sqrt{ 1 - f(t^m, y^m, v^m, w^m)}} 
\ee
up to an overall constant factor. 
Employing the method given in \cite{Kinney:2005ej} 
the index of single gauge-invariant operators is determined as 
\be 
I^{\cN=4}_{SO(\infty)} = -{1\over 4} + \frac{(1 - f(t, y, v, w))^2}{4 (1 - f(t^2, y^2, v^2, w^2))} + \half I^{\cN=4}_{U(\infty)}
\label{solargeNindex}
\ee
where $I^{\cN=4}_{U(\infty)}$ is the large $N$ limit of single gauge-invariant operator index of $\cN=4$ SYM with $U(N)$ gauge group given by 
\be 
I^{\cN=4}_{U(\infty)}= {vt^2 \over 1- vt^2} +{t^2/w \over 1- t^2/w} +{wt^2/v \over 1- wt^2/v} -{yt^3 \over 1- yt^3} - {t^3/y \over 1- t^3/y}.
\ee
Note that the constant term in \eqref{solargeNindex} is determined so that  vacuum contribution vanishes.  
This large $N$ index of the orthogonal gauge group \eqref{solargeNindex} exactly agrees with the single particle index describing the supergravity multiplet  
in $\bm{AdS}_5\times \RP^5$ given by $I^{\rm KK}_{\ZZ_2}$ by setting
\be
p= \frac{1}{\sqrt{w}},\quad q= \frac{\sqrt{w}}{v},\quad z= \frac{1}{v}.
\label{map}
\ee


\begin{thebibliography}{99}
\bibitem{Maldacena:1997re} 
  J.~M.~Maldacena,
  ``The Large N limit of superconformal field theories and supergravity,''
  Int.\ J.\ Theor.\ Phys.\  {\bf 38}, 1113 (1999)
  [Adv.\ Theor.\ Math.\ Phys.\  {\bf 2}, 231 (1998)]
  doi:10.1023/A:1026654312961
  [hep-th/9711200].
\bibitem{Ferrara:1998zt} 
  S.~Ferrara, M.~Porrati and A.~Zaffaroni,
  ``N=6 supergravity on AdS(5) and the SU(2,2/3) superconformal correspondence,''
  Lett.\ Math.\ Phys.\  {\bf 47}, 255 (1999)
  doi:10.1023/A:1007592711262
  [hep-th/9810063].
\bibitem{Beck:2016lwk}
  S.~W.~Beck, J.~B.~Gutowski and G.~Papadopoulos,
  ``AdS$_{5}$ backgrounds with 24 supersymmetries,''
  JHEP {\bf 1606}, 126 (2016)
  doi:10.1007/JHEP06(2016)126
  [arXiv:1601.06645 [hep-th]].
\bibitem{Garcia-Etxebarria:2015wns}
  I.~García-Etxebarria and D.~Regalado,
  ``$ \mathcal{N}=3 $ four dimensional field theories,''
  JHEP {\bf 1603}, 083 (2016)
  doi:10.1007/JHEP03(2016)083
  [arXiv:1512.06434 [hep-th]].
\bibitem{Aharony:2015oyb}
  O.~Aharony and M.~Evtikhiev,
  ``On four dimensional N = 3 superconformal theories,''
  JHEP {\bf 1604}, 040 (2016)
  doi:10.1007/JHEP04(2016)040
  [arXiv:1512.03524 [hep-th]].
\bibitem{Cordova:2016xhm} 
  C.~Cordova, T.~T.~Dumitrescu and K.~Intriligator,
  ``Deformations of Superconformal Theories,''
  arXiv:1602.01217 [hep-th].
\bibitem{Argyres:2016xua}
  P.~C.~Argyres, M.~Lotito, Y.~Lü and M.~Martone,
  ``Expanding the landscape of $ \mathcal{N} $ = 2 rank 1 SCFTs,''
  JHEP {\bf 1605}, 088 (2016)
  doi:10.1007/JHEP05(2016)088
  [arXiv:1602.02764 [hep-th]].
\bibitem{Nishinaka:2016hbw} 
  T.~Nishinaka and Y.~Tachikawa,
  ``On 4d rank-one $ \mathcal{N}=3 $ superconformal field theories,''
  JHEP {\bf 1609}, 116 (2016)
  doi:10.1007/JHEP09(2016)116
  [arXiv:1602.01503 [hep-th]].
\bibitem{Gunaydin:1984fk} 
  M.~Gunaydin and N.~Marcus,
  ``The Spectrum of the s**5 Compactification of the Chiral N=2, D=10 Supergravity and the Unitary Supermultiplets of U(2, 2/4),''
  Class.\ Quant.\ Grav.\  {\bf 2}, L11 (1985).
  doi:10.1088/0264-9381/2/2/001
\bibitem{Witten:1998xy} 
  E.~Witten,
  ``Baryons and branes in anti-de Sitter space,''
  JHEP {\bf 9807}, 006 (1998)
  doi:10.1088/1126-6708/1998/07/006
  [hep-th/9805112].
\bibitem{Aharony:2016kai}
  O.~Aharony and Y.~Tachikawa,
  ``S-folds and 4d N=3 superconformal field theories,''
  JHEP {\bf 1606}, 044 (2016)
  doi:10.1007/JHEP06(2016)044
  [arXiv:1602.08638 [hep-th]].
\bibitem{Schwarz:1983qr} 
  J.~H.~Schwarz,
  ``Covariant Field Equations of Chiral N=2 D=10 Supergravity,''
  Nucl.\ Phys.\ B {\bf 226}, 269 (1983).
  doi:10.1016/0550-3213(83)90192-X
\bibitem{Schwarz:1983wa} 
  J.~H.~Schwarz and P.~C.~West,
  ``Symmetries and Transformations of Chiral N=2 D=10 Supergravity,''
  Phys.\ Lett.\ B {\bf 126}, 301 (1983).
  doi:10.1016/0370-2693(83)90168-5
\bibitem{Kinney:2005ej} 
  J.~Kinney, J.~M.~Maldacena, S.~Minwalla and S.~Raju,
  ``An Index for 4 dimensional super conformal theories,''
  Commun.\ Math.\ Phys.\  {\bf 275}, 209 (2007)
  doi:10.1007/s00220-007-0258-7
  [hep-th/0510251].
\bibitem{McGreevy:2000cw} 
  J.~McGreevy, L.~Susskind and N.~Toumbas,
  ``Invasion of the giant gravitons from Anti-de Sitter space,''
  JHEP {\bf 0006}, 008 (2000)
  doi:10.1088/1126-6708/2000/06/008
  [hep-th/0003075].

\bibitem{Grisaru:2000zn} 
  M.~T.~Grisaru, R.~C.~Myers and O.~Tafjord,
  ``SUSY and goliath,''
  JHEP {\bf 0008}, 040 (2000)
  doi:10.1088/1126-6708/2000/08/040
  [hep-th/0008015].
\bibitem{Hashimoto:2000zp} 
  A.~Hashimoto, S.~Hirano and N.~Itzhaki,
  ``Large branes in AdS and their field theory dual,''
  JHEP {\bf 0008}, 051 (2000)
  doi:10.1088/1126-6708/2000/08/051
  [hep-th/0008016].

\bibitem{Balasubramanian:2001nh} 
  V.~Balasubramanian, M.~Berkooz, A.~Naqvi and M.~J.~Strassler,
  ``Giant gravitons in conformal field theory,''
  JHEP {\bf 0204}, 034 (2002)
  doi:10.1088/1126-6708/2002/04/034
  [hep-th/0107119].

\bibitem{Corley:2001zk} 
  S.~Corley, A.~Jevicki and S.~Ramgoolam,
  ``Exact correlators of giant gravitons from dual N=4 SYM theory,''
  Adv.\ Theor.\ Math.\ Phys.\  {\bf 5}, 809 (2002)
  [hep-th/0111222].

\bibitem{Corley:2002mj} 
  S.~Corley and S.~Ramgoolam,
  ``Finite factorization equations and sum rules for BPS correlators in N=4 SYM theory,''
  Nucl.\ Phys.\ B {\bf 641}, 131 (2002)
  doi:10.1016/S0550-3213(02)00573-4
  [hep-th/0205221].


\bibitem{Grant:2008sk} 
  L.~Grant, P.~A.~Grassi, S.~Kim and S.~Minwalla,
  ``Comments on 1/16 BPS Quantum States and Classical Configurations,''
  JHEP {\bf 0805}, 049 (2008)
  doi:10.1088/1126-6708/2008/05/049
  [arXiv:0803.4183 [hep-th]].
\bibitem{Yokoyama:2014qwa} 
  S.~Yokoyama,
  ``More on BPS States in $ \mathcal{N}=4 $ Supersymmetric Yang-Mills Theory on R $\times$ S$^{3}$,''
  JHEP {\bf 1412}, 163 (2014)
  doi:10.1007/JHEP12(2014)163
  [arXiv:1406.6694 [hep-th]].
\bibitem{Dolan:2008qi} 
  F.~A.~Dolan and H.~Osborn,
  ``Applications of the Superconformal Index for Protected Operators and q-Hypergeometric Identities to N=1 Dual Theories,''
  Nucl.\ Phys.\ B {\bf 818}, 137 (2009)
  doi:10.1016/j.nuclphysb.2009.01.028
  [arXiv:0801.4947 [hep-th]].

\bibitem{Spiridonov:2010qv} 
  V.~P.~Spiridonov and G.~S.~Vartanov,
  ``Superconformal indices of ${\mathcal N}=4$ SYM field theories,''
  Lett.\ Math.\ Phys.\  {\bf 100}, 97 (2012)
  doi:10.1007/s11005-011-0537-2
  [arXiv:1005.4196 [hep-th]].


\end{thebibliography}


\end{document}